\documentclass[conference]{IEEEtran}
\IEEEoverridecommandlockouts
\usepackage{cite}
\usepackage{amsmath,amssymb,amsfonts}
\usepackage{algorithmic}
\usepackage{graphicx}
\usepackage{textcomp}
\usepackage{multirow}
\usepackage{xcolor}
\def\BibTeX{{\rm B\kern-.05em{\sc i\kern-.025em b}\kern-.08em
    T\kern-.1667em\lower.7ex\hbox{E}\kern-.125emX}}
\begin{document}

\title{High-resolution home location prediction from tweets using deep learning with dynamic structure
\thanks{This material is based upon work supported by the National Science Foundation under Grant No. 1640822.}
}

\author{\IEEEauthorblockN{ Meysam Ghaffari}
\IEEEauthorblockA{\textit{Dept. of Computer Science} \\
\textit{Florida State University}\\
Tallahassee, US \\
ghaffari@cs.fsu.edu}
\and
\IEEEauthorblockN{Ashok Srinivasan}
\IEEEauthorblockA{\textit{Dept. of Computer Science} \\
\textit{University of West Florida}\\
Pensacola, US \\
asrinivasan@uwf.edu}
\and
\IEEEauthorblockN{Xiuwen Liu}
\IEEEauthorblockA{\textit{Dept. of Computer Science} \\
\textit{Florida State University}\\
Tallahassee, US \\
liux@cs.fsu.edu}
\and

}

\pagestyle{plain}

\maketitle

\begin{abstract}
Timely and high-resolution estimates of the home locations of a sufficiently large subset of the population are critical for effective disaster response and public health intervention, but this is still an open problem. Conventional data sources, such as census and surveys, have a substantial time lag and cannot capture seasonal trends. Recently, social media data has been exploited to address this problem by leveraging its large user-base and real-time nature. However, inherent sparsity and noise, along with large estimation uncertainty in home locations, have limited their effectiveness. Consequently, much of previous research has aimed only at a coarse spatial resolution, with accuracy being limited for high-resolution methods. {\em In this paper, we develop a deep-learning solution that uses a two-phase dynamic structure to deal with sparse and noisy social media data.} In the first phase, high recall is achieved using a random forest, producing more balanced home location candidates. Then two deep neural networks are used to detect home locations with high accuracy. We obtained over 90\% accuracy for large subsets on a commonly used dataset. Compared to other high-resolution methods, our approach yields up to 60\% error reduction by reducing high-resolution home prediction error from over 21\% to less than 8\%. Systematic comparisons show that {\em our method gives the highest accuracy both for the entire sample and for subsets}. Evaluation on a real-world public health problem further validates the effectiveness of our approach.

\end{abstract}


\section{Introduction}
Applications in diverse domains, including agriculture, transportation, poverty reduction, conflict prevention, disaster response, and humanitarian aid, require knowledge of the distribution of home locations of the population, or of specific demographic sub-groups, for effective public policy interventions~\cite{b1,b2,b3}. The conventional approach in these fields is to use census data or data from surveys, such as the American Community Survey (ACS). However, these are conducted too infrequently to provide timely information. Moreover, ACS data has a resolution of a zone containing 100,000 people, which is too coarse for critical applications explained later. 

New data sources, such as cell phone data records and GPS information, have been considered as alternative sources~\cite{b1, b2}. However, the use of cell phone data has strict regulatory constraints and is not widely accessible. Moreover, its granularity is limited by the closest base transceiver station antennas to the user.

Social media data can potentially address the spatial and temporal challenges. Social media activities often use the device GPS to provide geotags with high-accuracy as metadata. Furthermore, social media has wide popularity. For example, Twitter has over 300 million active users worldwide. The use of social media is also increasing, with the number of tweets per day at 500 million in June 2018, in contrast to 400 million in March 2013~\cite{b4}. This offers the potential of obtaining real-time information on a large population sample with high spatial resolution.

 These observations motivate the problem addressed in this paper. \textit{Given metadata for a large number of tweets, we wish to find home locations with 100m resolution for a subset of users, with high accuracy in the prediction}. 

Note that the American Community Survey, conducted by the US Census Bureau, has a sample size of around 2 million for each of the last few years. With approximately 67 million active Twitter users in the US, the ability to predict accurately for even 10\% of the users would provide us with a sample that is several times that of the American Community Survey. Besides, Twitter would deliver results in real-time, in contrast to the annual reports published by the latter.

Despite the promise of Twitter data, there are also significant challenges arising from incorrect, imprecise, or missing information. In particular, the home location in the Twitter profile is optional. Hecht et al. have determined that only $42\%$ of the Twitter users report a valid city on their Twitter profile \cite{b22}. Furthermore, users often provide home location at the city level, which is not sufficiently precise for the class of applications that we consider. As an alternative, others have considered inferring home location from users’ check-in activities using the geotags of tweets. The challenge here is that users tweet at multiple locations, which makes it hard to pinpoint the precise user home location out of several locations that they may visit.

Given the above challenges, much of prior research has focused on predicting home location at the state and city levels. The few papers in the literature dealing with high-resolution predictions use Support Vector Machines (SVM) with a linear kernel for prediction, obtaining 70\% accuracy for a 76\% subset of the test population with 100m resolution. The highly imbalanced and complex nature of the data limits the efficacy of such an approach. 

In this paper, we use a two-phase dynamic structure to manage the highly imbalanced and complex data effectively. In the first phase, we use a random forest designed to yield high recall to produce a more balanced set of records containing home candidates. In the second phase, using the more balanced sample, we train two different deep neural network models: Deep Neural Network for Regression (DNN-R) and Deep Neural Network for Classification (DNN-C). DNN-R is responsible for detecting user home location among available location records, and DNN-C is used for either approving or rejecting the detected record as the user home location, thus controlling the subset of data for which we provide a predicted home location.

By using a fast random forest algorithm to remove the majority of non-home records and then using more precise methods to improve accuracy, our approach yields the highest accuracy for high-resolution home location prediction from Twitter data for both the entire sample and for its subsets, obtaining up to 92.6\% for a 10\% subset and achieving up to 60\% prediction error reduction in comparison to  other methods.
In addition, as an application, we used the proposed method to detect high-risk neighborhoods for the 2016 Zika epidemic importation from Puerto Rico to Florida and showed that it was substantially more effective than conventional ACS data~\cite{b36}. 

The rest of the paper is organized as follows. We discuss related work on home location prediction and its applications in Section 2. We then describe our deep learning model in Section 3 and analyze its performance empirically in Section 4. We also demonstrate its effectiveness in detecting high-risk neighborhoods for Zika importation, compared with ACS data. We summarize our conclusions in Section 5 and present directions for future work.

\section{Related Work}
As mentioned above, several applications need high-resolution home location~\cite{b1, b2, b3}. The motivating application for us is the spread of vector-borne diseases such as Zika, Dengue, and Chikungunya. These are spread by mosquitoes of the Aedes genus which have a range of 100m-200m. If one can identify the home locations of people who recently visited regions experiencing an outbreak at such a granularity, then mosquito control measures can be cost-effectively deployed in those locations to reduce the likelihood of local disease spread \cite{b30}. It is sufficient for the mathematical models in these applications if we generate home locations of a sample of the population, provided it is large enough to capture the distribution of demographic groups of interest. As long as the sample size is sufficiently large, the primary goal is maximizing the accuracy \cite{b23}. 

Cell phone call data records (CDR) and GPS data \cite{b1},\cite{b12} have been  used for home location prediction. However, they are not widely used due to limited resolution, regulatory constraints, and their high cost. In comparison, there had been much recent interest in leveraging the abundance of social media data to predict users' home locations. For example, Backstrom et al. used Facebook users' friends to predict their home locations with an accuracy of $69.1\%$ within a range of 25 miles [15].  Most research related to home-location predictions from tweets has focused on coarse spatial scales, such as time zone, state, and city. Mahmud et al. predicted user home location based on tweet geotags at the city, state and time zone levels with the accuracies of $58\%$, $66\%$ and $78\%$ respectively \cite{b25}. Cheng et al. predicted the home location of users within 100 miles of their home with $51\%$ accuracy \cite{b26}. Pontes et al. also used Twitter geotags to detect user home locations at the city level with an accuracy of $82\%$ \cite{b6}. However, few researchers tried to predict at the fine-granularity that we target, until recently. 

In recent work, Tasse et al. focused on predicting home locations at finer spatial scales than in prior work \cite{b17}. They predicted the home location with a resolution of 1 KM with $79\%$ accuracy and within 100m with $56\%$ accuracy. Hu et al. extracted a few features for users check-in patterns and improved the accuracy of home location prediction to $70\%$ for a $76\%$ subset of the data using a Support Vector Classifier (SVC) with linear regression \cite{b20}. Kavak et al. defined two additional features for users' check-in patterns. They applied DBSCAN -- a density-based clustering algorithm -- to extract tweet locations for each user. Tweets with spatially close geotags from the same user are assigned to the same cluster~\cite{b39}. A feature vector corresponds to each cluster as shown in Table~\ref{tab:features} later. They then applied SVM with a linear kernel to train and test the model using 5-fold cross-validation. They achieved a best result of $79.5\%$ for predicting users home location within the range of 100m from their home \cite{b10}. Since this is the best-reported accuracy, and they made their dataset publicly available, we evaluated our work too on this dataset.

\section{Proposed Method}
\begin{figure}
\centering
\includegraphics[width=0.9\linewidth]{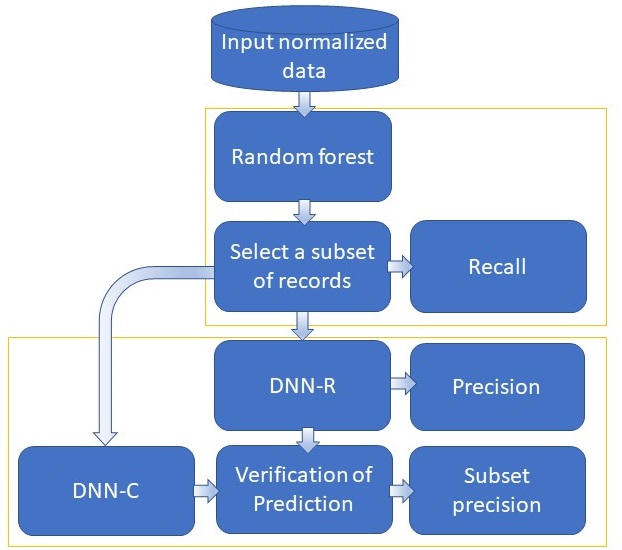}
\caption{The Proposed Method Workflow}
\label{fig:ijcnn}
\vspace{-0.225in}
\end{figure}

Tweets by a user may be from several locations. In our approach, we select one record which indicates the user's home location out of multiple records that indicate places that a user visited. In this problem, the user's home record is a minor class and other places the user visited is the major class. Disparity in the sizes of the two classes makes the problem unbalanced, which is exacerbated by the complexity of the data arising from travel pattern variations in the Twitter geotag dataset \cite{b33}. Dynamic structures are useful in problems with complex, unbalanced datasets, where it is not easy to detect the minority class \cite{b29}. We use a two-phase model. The first phase uses a simple algorithm with high recall, which runs fast and eliminates a significant number of records in the majority class to make the data for the next phase more balanced. The second phase, involves an effective but time-consuming algorithm to detect the minority class records precisely. Figure \ref{fig:ijcnn} shows the main components of the proposed method.

As shown in Figure 1, we first normalize the features. We then divide the dataset into a training set and a test set, and we train a random forest on the training set so that it can predict on the test set. We used 5 fold cross-validation on each time $80\%$ of the data is used for the training and $20\%$ for the test. After performing this phase with high recall, we obtain a significantly smaller dataset. The high recall ensures that we don't miss many true  home locations, while eliminating several records that are not the home. In the second phase, we again divide the newly created smaller dataset into training and test sets and train two different DNN models on the training set. The first DNN is a regression model (DNN-R) and the second DNN is a classification model (DNN-C). After training, DNN-R selects one record for each user that corresponds to the predicted home. This result can be used to obtain the precision for the entire test population. However, our focus is on identifying home locations of a subset of the population with high accuracy.  We additionally use DNN-C to accomplish this as follows. DNN-R sends records detected as home locations to DNN-C. DNN-C classifies each record as the home with a certain probability. If this probability exceeds a threshold, then the record will be reported as a user's home. Otherwise, that user's home will be reported as unknown. We provide below further details of each step.

\subsection{Feature Normalization}
Feature normalization is a standard step that ensures that all features are considered equally in the learning algorithm. It is accomplished using $ X'= 2 \times \frac{X-X_{min}}{X_{max}-X_{min}}-1 $ of each feature X, where $X_{min}$ and $X_{max}$ are the minimum and maximum values respectively of that feature, and X' is the normalized value. Feature normalization helps the algorithm to better model the dataset, and prevents bias toward one feature with high values. After normalization, we perform the first phase of our learning algorithm using a random forest.

\subsection{Random Forest}
Random forests are based on decision trees. A decision tree is a supervised classifier which gets a set of data and creates a tree-like model of rules for classifying the data. The rules are based on the features of the training data \cite{b37}. In a random forest, instead of creating one tree to classify the dataset, several trees are created. Each tree uses just a subset of features, and also a subset of training data is used for training the model. In typical use, the majority of trees define the class of each record. But we differ from this as explained later. The biggest benefit of random forest over the decision trees is that it works on a bootstrap dataset with randomly selected features for each tree, and thus tries to prevent overfitting. The bootstrap dataset is created using randomly selected records of the dataset with replacement, so that it contains the same number of records as the original dataset. Some records may be selected multiple times due to this form of selection. The random forest has an error rate comparable to AdaBoost \cite{b34}, but at the same time is more robust with respect to  noise. Moreover, it is proven to work well on imbalanced data \cite{b9}.

In the first phase of our model, we aim to eliminate non-home location records as much as possible. For this purpose, we use a random forest to classify the records as home or not home records. In order to have a high recall, we select every record that any tree in the forest predicted as the home location, rather than use the typical majority decision. We send the selected records in this phase to the second phase.

\subsection{Deep Neural Network for Regression (DNN-R)}

In the second phase of the algorithm, we designed and applied a deep learning model -- a multi layer perceptron. The configuration of the sequential fully connected deep neural network that we used is shown in Table \ref{tab:config1}. In this model, five dense layers have been used and input data has 10 features. In order to include non-linearity in the model, we used Rectified Linear Units (ReLU) activation functions in the first 4 layers \cite{b11} and sigmoid in the last layer. 

\begin{table}[]
\caption{The configuration of DNN-R and DNN-C}
\begin{center}
\begin{tabular}{|p{3.5cm}|p{3.5cm}|}
\hline
   DNN-R Configuration & DNN-C Configuration\\
   \hline
    Input dimension: 10 &   Input dimension: 10 \\
   \hline
   Dense1:  Input: 10, output: 5, Activation: ReLU &  Dense1:  Input: 10, output: 5, Activation: ReLU\\
  \hline
  
  Dropout1: 0.30 dropout rate & Dropout1: 0.20 dropout rate \\
  \hline
  Dense2:  output: 20, Activation: ReLU  & Dense2:  output: 20, Activation: ReLU \\
  \hline
  Dropout2: 0.30 dropout rate & Dropout2: 0.20 dropout rate \\
  \hline
  Dense3:  Input: 20, output: 5, Activation: ReLU &  Dense3:  Input: 20, output: 5, Activation: ReLU \\
  \hline
  Dropout3: 0.30 dropout rate &  Dropout3: 0.20 dropout rate \\
  \hline
  Dense4:  output: 5, Activation: ReLU & Dense4:  output: 5, Activation: ReLU \\
  \hline
  Dropout4: 0.30 dropout rate  & Dropout4: 0.20 dropout rate \\
  \hline
  Dense5: Output: 1, Activation: Sigmoid  &  Dense5: Output: 2, Activation: Sigmoid  \\       
  \hline

\end{tabular}
 \end{center}
 \label{tab:config1}
\end{table} 

 In order to prevent overfitting and to improve the generalization in the model, we applied dropout layers after each of the first four dense layers. The dropout randomly changes the weight of some neurons with the predefined probability to 0, thus preventing overfitting \cite{b31}. We used a regression model with a Stochastic Gradient Descent (SGD) optimizer and mean square error loss function. SGD is an iterative method for finding the optimum point of differentiable functions. Then we trained this model on the selected records in the first phase to detect the user home location where the home has the value of 1 and other records have the value of 0 in the target values. After training this model on the training set, we used it on the test set. For each user, the record with the highest prediction value is considered as the home location. 
 
 \subsection{Deep Neural Network for Classification (DNN-C)}
 
 The second deep learning model has a similar configuration with two differences, as seen from Table \ref{tab:config1}. Instead of regression, it is designed for classification. So, we used the categorical cross-entropy as the loss function and the ‘RMSprop’ optimizer \cite{b35}. For each weight, this optimizer divides the learning rate by considering a running average of the magnitudes for the recent gradients pertinent to that weight \cite{b32}. Furthermore, since the algorithm is a categorization algorithm having two classes, the last layer has two outputs for two different classes. 
 
In the last phase, by comparing the prediction value of DNN-C with a threshold, we verify the results of the DNN-R. The result will be reported only when the prediction value is higher than the threshold. Thus, instead of predicting the home location of all users, we predict the home location for a subset of users, but with higher accuracy.

\section{Results and Analysis}
\subsection{Description of Data Set}
We used a well-curated dataset prepared by Kavak et al. \cite{b10}. Their data was gathered using Twitter streaming API from May 2014 to April 2015 for the city of Chicago, Illinois. They performed anonymization to preserve the privacy of the users and then ran DBSCAN to cluster together tweets that are in close proximity, with the distance range specified as 100m. Each record in the final database relates to tweets from a particular user at a particular location with a 100m spatial resolution. For validation purposes, the true home location was determined by obtaining confirmation from the users about whether they tweeted from home. The final dataset has 78,812 records for 1268 users. The features of this dataset are listed in Table \ref{tab:features} \cite{b10}. 

 \begin{table}[]
 \caption{Dataset description}
 \begin{center}
 \begin{tabular}{|p{2.3cm}|p{4.7cm}|}
 \hline
    
    Feature & Description \\
    \hline
    
    Check-in ratio & The ratio of the number of check-ins in a specific location by a user to the total number of check-ins at all locations by that user. \\
    \hline    
    Daily total check-in rate & The average daily number of check-ins by the user. \\
   \hline

   End of day ratio & The ratio of the number of last check-in between 5PM-3AM of the day at a specific location to the same for all locations. \\
   \hline
   
   End of inactive day ratio & The ratio of the number of last check-in between 5PM-3AM of each weekend day at a specific location to the same for all locations.\\
   \hline
  
  Distance from most check-in location & The distance of a specific location from the most visited location by that user.\\
   \hline
  
  Midnight ratio & The ratio of the number of check-ins at a specific location between 12AM-7AM by a user to all check-ins during 12AM-7AM that user. \\
   \hline
  
  Number of check-ins at this location &Number of check-ins at this location by the user.\\
   \hline
   
   Total number of user check-ins & Total number of check-ins by the user.\\
   \hline
  
  Page rank & A graph measure to show the importance of each location. A node in a graph represents a location and the weight of a directed edge from u to v gives the number of times a user went from location u to location v. This measure considers the consequence of visited locations until 3AM of each day.\\
   \hline
   
   Reverse page rank & Is similar to page rank, but swapping the source and destinations.\\       
   \hline  
   
   User-ID & The unique ID assigned to each user in order to preserve privacy. \\
   \hline

   Is-home & Whether or not a record corresponds to the user's home. \\
   \hline

 \end{tabular}
  \end{center}
  \label{tab:features}
 \end{table} 

\subsection{Experimental Setup}
We used an identical dataset and test procedure as~\cite{b10} in order to ensure a fair comparison. We used 5-fold cross-validation in both phases of our model identical as state of the art, which means that the model is trained using $80\%$ of the dataset and validated using the remaining $20\%$ in each of five experiments. Both the DNN phases use the same training data in each test, and similarly the same test data. Note that records are selected into these sets based on the user. Consequently, all records for a specific user will either go into the training set or into the test set in any single experiment.

\subsection{Results of First Phase (Random Forest)}
We use a random forest  with 500 trees. The random forest predicts each record as user's home location with a prediction value between 0 and 1. This prediction value is the fraction of decision trees that considered that record as a user's home. We can select a record as a candidate for a user's home location if its prediction value exceeds a predefined threshold. In order to have high recall, we selected threshold as 0.002 in the forest with 500 trees; so, even if one tree in the forest predicts the record as home we select it for the next phase.

\begin{figure}
\centering
\includegraphics[width=0.9\linewidth]{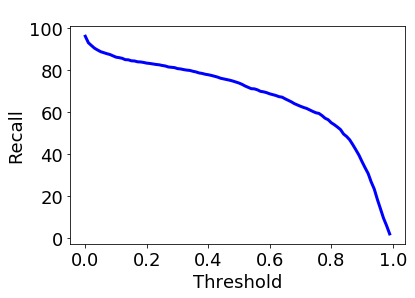}
\caption{Recall versus threshold in the output of the random forest}
\label{fig:Recall1}
\vspace{-0.175in}
\end{figure}

\begin{figure}
\centering
\includegraphics[width=1.1\linewidth]{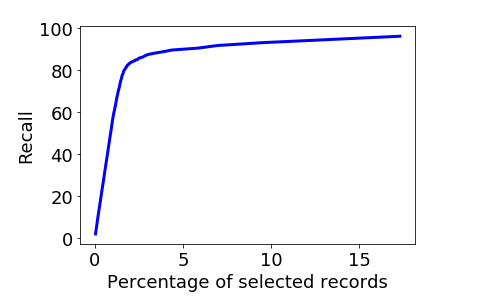}
\caption{Recall versus percentage of selected records for the output of the random forest}
\label{fig:Recall2}
\end{figure}

As shown in Figures \ref{fig:Recall1} and \ref{fig:Recall2}, selecting a higher threshold significantly decreases the recall, especially when the threshold is close to 1. Since we are looking for high recall, while also pruning the records that are clearly not a user's home location, we considered a threshold close to the zero. As shown in Figure 3, the highest recall is obtained by having $17.5\%$ of the records for the second phase and selecting fewer records will decrease the recall, which is not favorable to our goal. We used a small threshold, with a record being selected if even one tree selected it. This corresponds to a threshold of 0.002 in a forest with 500 trees. This yields a recall of $95.97\%$ with an average of 10.7 records per user, which is a substantial reduction over the roughly 62 records per user in the initial dataset. The selected records in this phase are sent to the second phase.

\subsection{Configuring the DNNs}

\begin{figure}
\centering
\includegraphics[width=.9\linewidth]{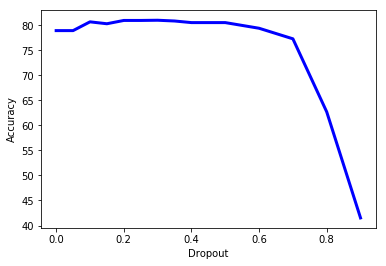}
\vspace{-0.2500in}
\caption{Effect of the value of the dropout}
\label{fig:Dropout}
\vspace{-0.200in}
\end{figure}

One of the important configurations of the DNN is the value of the dropout. As mentioned earlier, dropout is a useful technique for generalization. In order to find the best value for the dropout, we checked different values of dropout on the DNN-R. Figure \ref{fig:Dropout} Shows the effect of the dropout on the result. Based on this figure, we chose $0.30\%$ for the dropout in the DNN-R. 

 As we can see in this figure, increasing the value of the dropout at the beginning improves the results, and then make it worse. The reason is that using small value of dropout such as $0.20$ or $0.30$ prevents the algorithm from overfitting. But if we use a high value for the dropout such as $0.90$, it makes the weight of the majority of the neurons to 0 which does not let algorithm to learn any pattern and makes it work randomly.

\begin{figure}
\centering
\includegraphics[width=.9\linewidth]{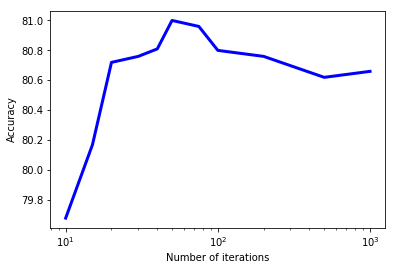}
\vspace{-0.15in}
\caption{Effect of number of epochs on the accuracy}
\label{fig:Iterations}
\vspace{-0.175in}
\end{figure}

Choosing an appropriate number of iterations for SGD is important in deep learning. If the number of iterations is too low, then it prevents the algorithm from fitting the data well and learning the pattern, while a high number of iterations can cause overfitting and decrease the generalization. We checked different numbers of iterations to find the best number for the algorithm. Figure \ref{fig:Iterations} shows the result of different number of iterations. Beyond 20 iterations, the results are not very sensitive to the number of iterations, with best results for 50 iterations. We used 50 iterations in our experiments.

\subsection{Results of the Second Phase (DNN-R and DNN-C)}
In the second phase, we have two types of results. First, the reported results based on  DNN-R show the accuracy of home location prediction for all users in the dataset. Second, the results of DNN-R combined with DNN-C which show the accuracy of home location prediction for a subset of users but with higher accuracy. 

After training DNN-R we apply it to the test set. For each user in the test set, we consider the record with the highest predicted value of the DNN-R model as the user home location. The results of predicting the home location of users are depicted in Table 5 and compared with the state of the art results. We can see that our model improves on the prediction for the whole population in the test set over prior methods. However, our primary goal is to obtain high accuracy in a subset of the population, which requires one more step, as described below.

In the final step, we use the record selected as a user's home location by DNN-R, and confirm whether this is true using DNN-C. DNN-C will provide a predicted score for each record provided to it. If this score exceeds a predefined threshold, then we will report that record as the user's home. Otherwise, we will report that user's home as unknown. Table \ref{tab:comparison} shows that this approach increased the home location prediction accuracy up to $92.6\%$ on a subset of users, which is significantly higher than the results of DNN-R for the entire test population, which averaged $81\%$ over the 5 tests, with a maximum of $83.4\%$ on one of those tests. Furthermore, this accuracy substantially exceeds those for prior results on subsets of the population. This accuracy is remarkable if we consider that Tasse et al. found that over $10\%$ of Twitter users did not have tweets within a range of 100m of their home location \cite{b17}. Consequently, the maximum possible accuracy for the entire population is less than $90\%$.

 \begin{table}[]
 \caption{Comparison of the new method and state of the art methods}
 \begin{center}
 \begin{tabular}{|p{2cm}|p{2cm}|p{3.2cm}|}
 \hline

    Method & Reported accuracy (100 meters resolution) & 	Description $\%$ \\
    \hline
        
   Hu et al. \cite{b20}: SVM & $70.00$ & For a subset of $76\%$ and $71\%$ of two different datasets \\
    \hline    
   Kavak et al. \cite{b10}: using DBSCAN and SVM  & $79.50$ & Best reported accuracy on the whole population among prior methods. They use the same dataset as we do. \\
   \hline

  Tasse et al. \cite{b17}: using multilevel DBSCAN and Grid search & $56.90$ & Reported result for 100 meter
  They also got $79\%$ for 1 KM resolution
  
   \\
   \hline
   
   Our model (DNN-R) & $\textbf{83.40}$ & Best achieved accuracy for the whole dataset\\
   \hline
  
  Our model (DNN-R + DNN-C) & $\textbf{85.10}$ &  Reported results for $80\%$ of the users \\
   \hline
  
 Our model (DNN-R + DNN-C) & $\textbf{91.86}$ & Reported results for $30\%$ of the users \\
   \hline
   
   Our model (DNN-R + DNN-C) & $\textbf{92.60}$ & Reported results for $10\%$ of the users \\
   \hline

 \end{tabular}
  \end{center}
  \label{tab:comparison}
  \vspace{-0.225in}
 \end{table} 

\begin{figure}
\centering
\includegraphics[width=.8\linewidth]{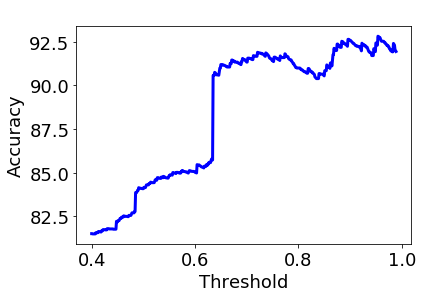}
\caption{The effect of threshold on the accuracy of the final results}
\label{fig:Threshold}
\end{figure}

\begin{figure}
\centering
\includegraphics[width=.8\linewidth]{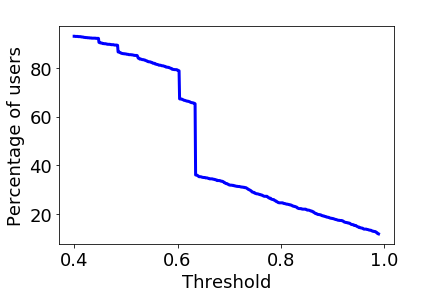}
\vspace{-0.15in}
\caption{Effect of threshold on the percentage of users with reported home location}
\label{fig:Threshold2}
\vspace{-0.225in}
\end{figure}

\begin{figure}
\centering
\includegraphics[width=.9\linewidth]{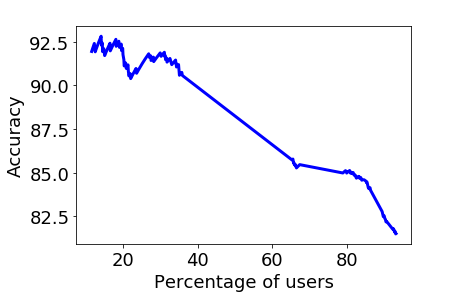}
\vspace{-0.225in}
\caption{The accuracy versus the percentage of users with reported home location by the method}
\label{fig:accuracy}
\vspace{-0.225in}
\end{figure}

Figures \ref{fig:Threshold},\ref{fig:Threshold2}, and \ref{fig:accuracy} shows the effect of the threshold on the accuracy and fraction of the population for whom our model can predict the home location. There is a trade-off between accuracy and the fraction of population for whom we can predict the home location. One can obtain the highest possible accuracy of around $92.6\%$ using $10\%$ of the total population. However, we can obtain a much larger sample -- $30\%$ -- without a substantial drop in accuracy, maintaining it at over $90\%$. Given the large number of Twitter users, our method can yield a large sample with good accuracy. 

As shown in the Figure \ref{fig:Threshold}, the accuracy increase with increase in the threshold. But the rate is not fixed. The reason is that by increasing the threshold in DNN-C, more records will be pruned. These records can belong into both true classified and wrong classified and their distribution is not the same. This means each group can have different records with different prediction value using DNN-C. However, in general it is effective and will increase the accuracy up to more than $92.6\%$. 

The same happens in Figure \ref{fig:accuracy}. When we prune more records, we will have higher accuracy for a smaller portion of the users. In this figure, we have the accuracy of $92.6\%$ but for $10\%$ of the users. If we want to report more users home location, we should decrease the threshold and this will lead to increasing percentage of users and decreasing the accuracy for them.

We expect our method to also be effective on other datasets because we have taken steps to ensure robustness, to avoid overfitting. For example, we used dropout in the deep neural networks. Moreover, we used 5-fold cross validation and performed the experiments at least 25 times. Thus, the whole dataset has been tested 5 times. The reported results are the average accuracy from these tests. The individual result did not vary much, and so the average is a reasonable reflection of typical performance.

\subsection{Analysis of Each Component}
We next analyze the contribution of specific model components to accuracy and training time. Table \ref{tab:separatetests} shows the results of the random forest and DNN-R. As we can see, using the random forest in the first phase decreases the execution time and also improves the final result as explained. We can also see that the result of each separate component is not as good as the mixed ones, which shows us that the dynamic structure is effective. In particular, the random forest decreases the training time and the balanced data that it produces enhances the effectiveness of the remaining components.

 \begin{table}[]
 \caption{The results of running random forest of DNN-R separately}
 \begin{center}
 \begin{tabular}{|p{2cm}|p{2cm}|p{2cm}|}
 \hline

    Method & Reported accuracy (100 meters resolution) & training time \\
    \hline
        
  Random Forest & $79.38\%$ & 719s \\
    \hline    
  DNN-R & $80.04\%$ & 9,485s \\
   \hline
   DNN-C & $72\%$ & 54,000s \\
   \hline
   DNN-R and DNN-C & $80.21\% - 83.82\%$ & 63,485s \\
   \hline
    Random forest, DNN-R and DNN-C & $84\% - 92.6\%$ & 26244s \\
    \hline
   \end{tabular}
  \end{center}
  \label{tab:separatetests}
  \vspace{-0.225in}
 \end{table} 

\subsection{Results in Detecting High Risk Neighborhoods in Zika 2016}
Here, we discuss the application of our method on a public health problem. We show that our method yields better results than use of conventional data sources for this application. In 2016, a Zika virus outbreak in Florida occurred through importation from the Caribbean and South America, with Puerto Rico playing a major role. Consequently, several Zika cases had been reported in Miami. They were mostly imported, though there was subsequent local spread. In 2016, CDC announced three Zika red zones of the order of a square mile each in Miami, which were Miami Beach, Wynwood, and Little River\cite{b36}. The conventional approach to identifying high-risk neighborhoods before an outbreak would detect places in Miami where persons with a connection to Puerto Rico lived. This relies on the assumption that such individuals were more likely than the general population to have visited Puerto Rico, which was experiencing a major outbreak, and been exposed to the virus there. ACS data provides the number of persons with origin in Puerto Ricans living in each Public Use Microdata Areas code (PUMA). Table \ref{tab:ACS} shows this based on 2016 data. As we can see, the red zones cannot be easily identified from this data.

 \begin{table}[]
 \caption{Number of first and second generation Puerto Ricans living in each PUMA zone. The Red Zones are in bold.}
 \begin{center}
 \begin{tabular}{|p{0.9cm}|p{2cm}|p{3cm}|p{0.7cm}|}
 \hline

    PUMA code & Close Neighborhood & Puerto Rican Generations: First + Second & Sum\\
    \hline
    08607 & North East Airport & 2814 + 1322 & 4136 \\
 \hline
  08601	& Miami Lakes	&3320 + 752 & 4072 \\
    \hline 
    
     08603 & North Miami Beach & 3470 + 495 & 3965
 \\
 \hline

     08605 & North Miami, Golden Glades & 3760 + 87 & 3847\\
 \hline
  08616 & West Miami & 1774 + 2069 & 3843\\
\hline
 
 08614 & Key Biscayne & 2623 + 926 & 3549 \\
\hline
08613 & Downtown & 2887 + 378 & 3274 \\
\hline

08611 & \textbf{Wynwood} & 3034 + 87 & 3121 \\
\hline
08610 & Miami Springs, Virginia Gardens & 2798 + 230 & 3028 \\
\hline
 08602	& Miami Gardens	& 2676 + 0 & 2676\\
   \hline
08615 & Coral Gables & 2018 + 394 & 2412 \\
\hline
 08606 & \textbf{West Little River} & 2054 + 305 & 2359 \\
 \hline
 08612 & \textbf{Miami Beach} & 1954 + 212 & 2166 \\
\hline
 08604 & North Beach, Bal Harbour & 2003 + 0 & 2003 \\
 \hline
08608 & Air Port & 1716 + 286 & 2002 \\
\hline
08609 & Hialeah & 785 + 485 & 1270\\
\hline

   \end{tabular}
  \end{center}
  \label{tab:ACS}
  \vspace{-0.25in}
 \end{table} 

Alternatively, we can use our algorithm to find high risk neighborhoods in Miami as follows. We extracted tweets of more than 500,000 users using Twitter API. We wished to generate a sample of users biased toward individuals in Florida with a connection to Puerto Rico. So, we started with few popular Twitter accounts in Puerto Rico and extracted their followers, the followers' followers, and so on, and kept those with profile data indicating location in Florida. We applied our model, trained on the earlier data set from Chicago, to find the home location of people who lived in Miami and had visited Puerto Rico. The results are shown in Table \ref{tab:neighborhoods}.

 \begin{table}[]
 \caption{Detected neighborhoods for Miami residents who visited Puerto Rico }
 \begin{center}
 \begin{tabular}{|p{3cm}|p{3cm}|}
 \hline

    Neighborhood & Percentage of users\\
    \hline
        
  Downtown & $25\%$ \\
    \hline    
 \textbf{Miami Beach} & $20\%$ \\
   \hline
   \textbf{Wynwood} & $10\%$ \\
 \hline
 Miami Airport & $10\%$\\
 \hline
 Allapattah & $10\%$ \\
 \hline

   \end{tabular}
  \end{center}
  \label{tab:neighborhoods}
  \vspace{-0.25in}
 \end{table} 

We detected five neighborhoods that could be at risk, which  included two of three red zones announced by CDC \cite{b36}. These neighborhoods are not among the neighborhoods with the highest Puerto Rican connection based on ACS data, which shows the efficiency of our new method and of social media data in identifying high-risk neighborhoods. Little River was a red zone that our method missed. Florida Department of Health confirmed to us, after we provided them our result, that this neighborhood did not have significant importation from Puerto Rico.

\section{CONCLUSION AND FUTURE WORKS}

In this paper, we focused on predicting the home location of subsets of Twitter users with high-resolution and high accuracy. We performed this task using a dynamic structure. A random forest was used to provide better balanced data. We then applied two different deep neural networks, one for prediction and the other one for validation. Using the DNN-R, we obtained an average $81\%$ accuracy for the whole population with the best result being over $83\%$, which is higher than the state of the art methods. More importantly, we obtained up to $92.6\%$ accuracy for a subset of Twitter users by using DNN-C to prune some of the results. This offers a variety of applications the option of obtaining real time home location data with fine spatial resolution. We then demonstrated the practical effectiveness of our method and of social media data by identifying neighborhoods at high risk of importing Zika from Puerto Rico in 2016. Our approach was much more effective than the conventional approach using ACS data.

One direction for future work is to use this technique in other applications mentioned earlier. Another direction is to increase the accuracy of the technique. For example, adding additional features could improve the accuracy of the algorithm. In addition, alternate algorithms for high recall in the first phase can impact the performance of the other phase. We will also explore improved machine learning approaches for the second phase.

\section*{Acknowledgement}
The authors thank Danielle Stanek and Andrea Morrison from the Florida Department of Health and Kelly Deutsch from the Orange County (Florida) Mosquito Control Division for information on the 2016 Zika outbreak and vector control challenges respectively.


\begin{thebibliography}{00}
\bibitem{b1} Jones KH, Daniels H, Heys S, Ford DV. Challenges and Potential Opportunities of Mobile Phone Call Detail Records in Health Research: Review. JMIR Mhealth Uhealth, 6:e161, 2018.
\bibitem{b2} Peak CM, Wesolowski A, Erbach-Schoenberg EZ, Tatem AJ, Wetter E, Lu X, Power D, Weidman-Grunewald E, Ramos S, Moritz S, Buckee CO, and Bengtsson L. Population mobility reductions associated with travel restrictions during the Ebola epidemic in Sierra Leone: use of mobile phone data. International Journal of Epidemiology, vol. 47, pp. 1562–1570, 2018.
\bibitem{b3} Wesolowski A, Qureshi T, Boni MF, Sundsoy PR, Johansson MA, Rasheed SB, Engo-Monsen K, and Buckee CO. Impact of human mobility on the emergence of dengue epidemics in Pakistan. Proceedings of the National Academy of Sciences, vol. 112, 11887-11892, 2015.
\bibitem{b4} Mahmud, Jalal, Jeffrey Nichols, and Clemens Drews. "Home location identification of twitter users." ACM Transactions on Intelligent Systems and Technology (TIST) 5.3 (2014): 47.

\bibitem{b5} https://www.omnicoreagency.com/twitter-statistics/   [26/10/18]
\bibitem{b6}Pontes, Tatiana, et al. "Beware of what you share: Inferring home location in social networks." Data Mining Workshops (ICDMW), 2012 IEEE 12th International Conference on. IEEE, 2012.

\bibitem{b7}Cho, Eunjoon, Seth A. Myers, and Jure Leskovec. "Friendship and mobility: user movement in location-based social networks." Proceedings of the 17th ACM SIGKDD international conference on Knowledge discovery and data mining. ACM, 2011.

\bibitem{b9}Chen, Chao, Andy Liaw, and Leo Breiman. "Using random forest to learn imbalanced data." University of California, Berkeley 110 (2004): 1-12.


\bibitem{b10}Kavak, Hamdi, Daniele Vernon-Bido, and Jose J. Padilla. "Fine-Scale Prediction of People’s Home Location using Social Media Footprints." International Conference on Social Computing, Behavioral-Cultural Modeling and Prediction and Behavior Representation in Modeling and Simulation. Springer, Cham, 2018.



\bibitem{b11} Liu, Ziyu, et al. "Top-Down Person Re-Identification With Siamese Convolutional Neural Networks." 2018 International Joint Conference on Neural Networks (IJCNN). IEEE, 2018.



\bibitem{b12} Isaacman, Sibren, et al. "Identifying important places in people’s lives from cellular network data." International Conference on Pervasive Computing. Springer, Berlin, Heidelberg, 2011.




\bibitem{b14}Zheng, Danning, et al. "Towards Lifestyle Understanding: Predicting Home and Vacation Locations from User's Online Photo Collections." ICWSM. 2015.



\bibitem{b15}Backstrom, Lars, Eric Sun, and Cameron Marlow. "Find me if you can: improving geographical prediction with social and spatial proximity." Proceedings of the 19th international conference on World wide web. ACM, 2010.

\bibitem{b16} Pontes, Tatiana, et al. "We know where you live: privacy characterization of foursquare behavior." Proceedings of the 2012 ACM conference on ubiquitous computing. ACM, 2012.

\bibitem{b17}Tasse, Dan, Alex Sciuto, and Jason I. Hong. "Our House, in the Middle of Our Tweets." ICWSM. 2016.

\bibitem{b18}Han, Bo, Paul Cook, and Timothy Baldwin. "Text-based twitter user geolocation prediction." Journal of Artificial Intelligence Research 49 (2014): 451-500.


\bibitem{b20}Hu, Tian-ran, et al. "Home location inference from sparse and noisy data: models and applications." Frontiers of Information Technology $\&$ Electronic Engineering 17.5 (2016): 389-402.

\bibitem{b22} Hecht, Brent, et al. "Tweets from Justin Bieber's heart: the dynamics of the location field in user profiles." Proceedings of the SIGCHI conference on human factors in computing systems. ACM, 2011.

\bibitem{b23}Ghaffari, Meysam, and Nasser Ghadiri. "Ambiguity-driven fuzzy C-means clustering: how to detect uncertain clustered records." Applied Intelligence 45.2 (2016): 293-304.


\bibitem{b24}Ajao, Oluwaseun, Jun Hong, and Weiru Liu. "A survey of location inference techniques on Twitter." Journal of Information Science 41.6 (2015): 855-864.


\bibitem{b25}Mahmud, Jalal, Jeffrey Nichols, and Clemens Drews. "Where Is This Tweet From? Inferring Home Locations of Twitter Users." ICWSM 12 (2012): 511-514.

\bibitem{b26}Cheng, Zhiyuan, James Caverlee, and Kyumin Lee. "You are where you tweet: a content-based approach to geo-locating twitter users." Proceedings of the 19th ACM international conference on Information and knowledge management. ACM, 2010.




\bibitem{b28} Zheng, Xin, Jialong Han, and Aixin Sun. "A survey of location prediction on Twitter." IEEE Transactions on Knowledge and Data Engineering (2018).


\bibitem{b29} Goodfellow, Ian, et al. Deep learning. Vol. 1. Cambridge: MIT press, 2016.


\bibitem{b30} Caminade C, Turner J, Metelmann S, Hesson JC, Blagrove MS, Solomon T, Morse AP, Baylis M. Global risk model for vector-borne transmission of Zika virus reveals the role of El Niño 2015. Proceedings of the National Academy of Sciences.  2017; 114(1), p. 119-124.

\bibitem{b31} Srivastava, Nitish, et al. "Dropout: a simple way to prevent neural networks from overfitting." The Journal of Machine Learning Research 15.1 (2014): 1929-1958.


\bibitem{b32} Chollet, Francois. Deep learning with python. Manning Publications Co., 2017.

\bibitem{b33} Hossain, Nabil, et al. "Precise Localization of Homes and Activities: Detecting Drinking-While-Tweeting Patterns in Communities." ICWSM. 2016.

\bibitem{b34} Rätsch, Gunnar, Takashi Onoda, and K-R. Müller. "Soft margins for AdaBoost." Machine learning 42.3 (2001): 287-320.

\bibitem{b35} Tieleman, Tijmen, and Geoffrey Hinton. "Lecture 6.5-rmsprop: Divide the gradient by a running average of its recent magnitude." COURSERA: Neural networks for machine learning 4.2 (2012): 26-31.

\bibitem{b36} https://www.cdc.gov/zika/intheus/florida-update.html.
\bibitem{b37} Safavian, S. Rasoul, and David Landgrebe. "A survey of decision tree classifier methodology." IEEE transactions on systems, man, and cybernetics 21.3 (1991): 660-674.

\bibitem{b38} Lathia, Neal, Daniele Quercia, and Jon Crowcroft. "The hidden image of the city: sensing community well-being from urban mobility." International conference on pervasive computing. Springer, Berlin, Heidelberg, 2012.

\bibitem{b39} Birant, Derya, and Alp Kut. "ST-DBSCAN: An algorithm for clustering spatial–temporal data." Data \& Knowledge Engineering 60.1 (2007): 208-221.

\end{thebibliography}
\end{document}